# Large family of two-dimensional ferroelectric metals discovered via machine learning

Xing-Yu Ma[1], Hou-Yi Lyu[1, 2], Kuan-Rong Hao[1], Yi-Ming Zhao[2], Xiaofeng Qian[3], Qing-Bo Yan[2*], Gang Su[4,1*]

[1]School of Physical Sciences, University of Chinese Academy of Sciences, Beijing 100049, China.

[2]Center of Materials Science and Optoelectronics Engineering, College of Materials Science and Optoelectronic Technology, University of Chinese Academy of Sciences, Beijing 100049, China.

[3]Department of Materials Science and Engineering, College of Engineering and College of Science, Texas A&M University, College Station, Texas 77843, USA.

[4]Kavli Institute for Theoretical Sciences, and CAS Center for Excellence in Topological Quantum Computation, University of Chinese Academy of Sciences, Beijing 100190, China.

*Correspondence authors.

*E-mail addresses:* yan@ucas.ac.cn (Q. B. Yan), gsu@ucas.ac.cn (G. Su).

ABSTRACT

Ferroelectricity and metallicity are usually believed not to coexist because conducting electrons would screen out static internal electric fields. In 1965, Anderson and Blount proposed the concept of "ferroelectric metal", however, it is only until recently that very rare ferroelectric metals were reported. Here, by combining high-throughput *ab initio* calculations and data-driven machine learning method with new electronic orbital based descriptors, we systematically investigated a large family (2,964) of two-dimensional (2D) bimetal phosphates, and discovered 60 stable ferroelectrics with out-of-plane polarization, including 16 ferroelectric metals and 44 ferroelectric semiconductors that contain seven multiferroics. The ferroelectricity origins from spontaneous symmetry breaking induced by the opposite displacements of bimetal atoms, and the full-*d*-orbital coinage metal elements cause larger displacements and polarization than other elements. For 2D ferroelectric metals, the odd electrons per unit cell without spin polarization may lead to a half-filled energy band around Fermi level and is responsible for the metallicity. It is revealed that the conducting electrons mainly move on a single-side surface of the 2D layer, while both the ionic and electric contributions to polarization come from the other side and are vertical to the above layer, thereby causing the coexistence of metallicity and ferroelectricity. Van der Waals heterostructures based on ferroelectric metals may enable the change of Schottky barrier height or the Schottky-Ohmic contact type and induce a dramatic change of their vertical transport properties. Our work greatly expands the family of 2D ferroelectric metals and will spur further exploration of 2D ferroelectric metals.

# 1. Introduction

Since the successful exfoliation of graphene [1] in 2004, numerous two-dimensional (2D) materials with extraordinary properties and rich potential applications have been discovered [2-4]. Ferroelectricity is an intriguing character of materials with switchable spontaneous electric polarization, which was generally believed to decay and even disappear if the film thickness is below a critical value [5-8]. For instance, $BaTiO_3$ thin films lose their ferroelectricity below a critical thickness of about six unit cells [9]. However, several 2D materials have recently been reported to be ferroelectrics, including Group IV monochalcogenides [10-12], 1T monolayer $MoS_2$ [13], buckled CrN and $CrB_2$ [14], $In_2Se_3$ and other $III_2$-$VI_3$ compounds [15,16], MXenes ($Sc_2CO_2$) [17], $AgBiP_2Se_6$, $CuMP_2X_6$ (M=Cr, V; X=S, Se) and $CuInP_2S_6$ [18-21], in contrast to the conventional notion that ferroelectricity would disappear in 2D limit [5-8], suggesting an underexplored exciting realm of 2D materials.

It is often thought that ferroelectricity and metallicity cannot coexist in a metal because conduction electrons would screen out static internal electric field that arises from dipole moment, thereby precluding intrinsic ferroelectric polarization [22,23]. In 1965, Anderson and Blount introduced the concept of 'ferroelectric metal' and proposed that the polar structure possibly appears in certain martensitic transitions involving the inversion symmetry breaking [24]. Recently, polar metals (e.g., $LiOsO_3$ and chemically tuned $MoTe_2$) have been reported [25,26]. Monolayer CrN and layered $Bi_5Ti_5O_{17}$ are also predicted to be ferroelectric metals [14,27]. More interestingly, two- or three-layer $WTe_2$ have been found to exhibit spontaneous electric polarization that can be switched using an external electric field [23], which may be the first experimental evidence for the coexistence of ferroelectricity and metallicity in a 2D material. Very recently, ferroelectricity driven nonlinear anomalous Hall current switching was proposed and the time-reversal invariance was experimentally demonstrated in odd-layer $WTe_2$ [28-30]. Nevertheless, until now, the examples of ferroelectric metals are still extremely sparse.

By combining high-throughput *ab initio* calculations and a data-driven machine learning model with new electronic orbital-based descriptors, here we systematically investigated a large family (2,964) of 2D bimetal phosphates, and discovered total 60 stable ferroelectrics, including 16 ferroelectric metals and 44 ferroelectric semiconductors among which seven multiferroics and seven ferroelectric water-splitting photocatalysts are screened out. The physical origin of ferroelectricity in these 16 2D ferroelectric metals is owing to the spontaneous symmetry breaking induced by the opposite vertical displacements of bimetal atoms. The ferroelectric-paraelectric transitions were simulated, revealing that the polarization could be reversed by a vertical external electric field. These ferroelectric metals possess odd electrons in a unit cell, in which conducting electrons mainly distribute on a single-side surface, while the ionic and electric contributions to polarization come from the other side, causing the coexistence of ferroelectricity and metallicity. The present work highly enriches the family of ferroelectric metals, suggesting that ferroelectric metals could be achievable in 2D materials. As ferroelectric metals could be constructed van der Waals heterostructures that may have wide applications in areas of ferroelectric tunneling junction, nonvolatile ferroelectric memory, etc., our proposal would spur great interest in exploring 2D ferroelectric metals in physics, materials sciences and information technology.

## 2. Materials and methods

The density functional theory first-principles calculations are performed by projected augmented wave (PAW) [31] implemented Vienna *ab initio* simulation package (VASP) [32]. The exchange-correlation interactions are treated using Perdew-Burke-Ernzerhof generalized gradient approximation (PBE-GGA) [33]. Cut-off energy of 450 eV was set for the plane-wave basis and $10 \times 10 \times 1$ k-points are used to sample the Brillouin zone. The convergence criteria were $1 \times 10^{-6}$ eV for the energy difference in the electronic self-consistent calculation and $1 \times 10^{-3}$ eV/Å for the residual forces on lattice geometries and atomic positions. All electronic structures are calculated using the PBE+$U$ method. For the on-site Coulomb interaction $U$ of the 3d, 4d and 5d transition metals, $U$ = 4, 2.5, and 0.5 eV are used, respectively, which are usually reasonable for them [34,35]. The all magnetic configurations are considered on $2 \times 2 \times 1$ unit cell. See the Supplementary materials for more details.

## 3. Results and Discussion

### *3.1 Structures of 2D bimetal phosphates*

Fig. 1a illustrates the schematic structure of 2D bimetal phosphates ($M_IM_{II}P_2X_6$, $M_I$ and $M_{II}$ atoms are different metal elements, X is chalcogen atom), which contains a honeycomb lattice (indicated by dash lines) formed by staggering metal atoms $M_I$ (blue balls) and $M_{II}$ (red balls), and P-P pairs (yellow balls) are located vertically at the center of hexagons. Chalcogen atoms (green balls) bridge metal atoms and P-P pairs, and each metal or phosphorous atom is surrounded by six or three chalcogen atoms, respectively. As indicated in Fig. 1b, if $M_I$ and $M_{II}$ atoms locating on the plane bisect perpendicularly the P-P pairs, the whole structure has a space group *P-6m2* (No.187) (or *P312* (No. 149), *P222* (No.16) for some materials), which corresponds to non-polar point groups *-6m2* (or *32*, *222*) and is denoted as "high-symmetry phase". Interestingly, among all 2D $M_IM_{II}P_2X_6$ materials, we found two types of possible spontaneous geometric symmetry reduction: (i) Type-I in Fig. 1c, $M_I$ and $M_{II}$ atoms deviate from the bisect plane (indicated with grey color) in opposite directions with displacements $d_1$ and $d_2$, respectively, and the total relative vertical displacement between $M_I$ and $M_{II}$ atoms is $d=d_1+d_2$. Meanwhile, the space groups reduce to *P3m1* (No. 156) (*or P3* (No. 143)*, P1* (No. 1)), which corresponds to polar point group *3m* (or *3*, *1*). (ii) Type-II in Fig. 1d, where the whole structure is distorted with P-P pairs inclining to three different directions, forming three symmetry-equivalent phases (*α*, *β*, and *γ*) with space groups *Cm* (No. 8) (or *P1*), which corresponds to polar point group *m* (or *1*). Our calculations reveal that the high-symmetry phase, Type-I and Type-II low-symmetry phases are paraelectric, ferroelectric, and ferroelastic (with ferroelectric), respectively.

Based on the above 2D bimetal phosphate prototype, we generate different structures of $M_IM_{II}P_2X_6$ by replacing $M_I$ and $M_{II}$ with 39 metal elements (Table S1 online) and replacing X by four chalcogen atoms (O, S, Se, and Te), respectively. The total number of such structures is 2,964, in which stoichiometrically equivalent structures are excluded, i.e., $M_IM_{II}P_2X_6$ and $M_{II}M_IP_2X_6$ are treated as the same material. A thorough first-principles investigation on such a large amount of materials would be extremely time-consuming. Here we use data-driven machine learning method by introducing new descriptors and combine with high-throughput *ab initio* calculations to accelerate the discovery of ferroelectrics from 2,964 structures of $M_IM_{II}P_2X_6$ materials.

### *3.2 Machine learning model*

The workflow is schematically illustrated in Fig. 2. Among 2,964 $M_IM_{II}P_2X_6$ structures, 605 of them are randomly selected as the initial training/test dataset for training a machine learning classification model. The structural optimization and corresponding properties were calculated by first-principles density functional theory (DFT) (the calculation details can be found in Materials and methods). Based on the results of these DFT calculations, 103 ferroelectrics (FE) were identified with a simplified criterion (Fig. S1 online). In the training of machine learning model, the materials in the dataset are described by 35 initial features (descriptors) (Table S2 online), including novel orbital-based descriptors designed by us, which were proved essential for a high-precision prediction (Table S3 online). Feature reduction was performed, and top 10 features were obtained to construct optimal feature space (Fig. S3 online). Five different machine learning algorithms such as the support vector classifier (SVC) [36], random forest classifier (RFC) [37], adaboost classifier [38], decision trees classifier (DTC) [39] and gradient boosting classifier (GBC) [40] were tested, all of which have been successfully applied to predict various materials [41,42]. The results of 5-fold cross-validation analysis and grid search for optimal hyper-parameters show that the GBC model outperforms the other four and gives the best performance (Fig. S2 online). Consequently, GBC was adopted in our model.

The precision of the initially obtained results was only 64%. To improve the performance, we introduced the data-driven methodology recently applied in materials and chemical sciences [43,44]. With the initial classification model, we obtained the prediction probability (Prob) of the remaining unexplored $M_IM_{II}P_2X_6$ structures and then labeled them as a positive or negative class with the criteria Prob ≥ 0.5 or Prob < 0.5, respectively. Those with the prediction probability near the dividing line, i.e., 0.45 ≤ Prob ≤ 0.55, were added to the training/test dataset. Based on the updated dataset, a new machine learning model could be obtained with improved precision. The above process was repeated until the model precision converges, which occurs at the fourth iteration (Figs. S5a online). Subsequently, a total of 293 extra $M_IM_{II}P_2X_6$ materials were added to the training/test dataset. In the end, we obtained an optimal machine learning model with high precision (77.2%) and high AUC (the area under the receiver operating characteristic curve) value (88.3%) (Fig. S5a and b online), showing that the data-driven methodology improved remarkably the performance of the machine-learning model. With this optimal classification model, 166 potential ferroelectrics are screened out from the remaining unexplored 2066 bimetal phosphates. Together with those 279 ferroelectrics identified in the updated training/test dataset, we obtained 445 potential ferroelectrics bimetal phosphates.

*3.3 Ferroelectrics*

As shown in the right panel of Fig. 2, we then performed systematic DFT calculations for these 445 ferroelectric candidates to acquire the optimized geometric structures with magnetic ground states. Their dynamical stabilities were examined by using density functional perturbation theory (DFPT) calculations [45], which yields 60 dynamically stable ferroelectric bimetal phosphates out of 445 candidates. In addition, their thermodynamic stabilities have been verified with the heat of formation (see details in the Supplementary materials). It should be mentioned that the energy above the convex hull can also be adopted to check the thermodynamic stability of a structure, which needs the information of its bulk phase [46]. As the present 2D structures have no corresponding bulk phases in databases, we opt to use the heat of formation to inspect their thermodynamic stabilities. They are displacive-type ferroelectrics, as their physical origin of ferroelectricity is owing to the spontaneous symmetry breaking mainly induced by vertical displacements of $M_I$ and $M_{II}$ atoms as shown in Fig. 1. For these 60 ferroelectric $M_I M_{II} P_2 X_6$ materials, we observed a quadratic relation between the out-of-plane polarization (***P***) and the relative vertical displacement (*d*) between metal atoms $M_I$ and $M_{II}$ (Fig. S6 online), i.e., $P \propto d^2$, providing a practical approach for rapidly identifying ferroelectric bimetal phosphates candidates with large electric polarization. It is quite different from the case of ferroelectric inorganic perovskites, in which the polarization is linearly correlated with the displacement [47].

The systematic DFT calculations on electronic structures of these 60 2D ferroelectric materials reveal that 16 of them are ferroelectric metals and the other 44 are ferroelectric semiconductors. This large family of novel ferroelectric metals will be discussed in detail below. Surprisingly, seven out of 44 stable ferroelectric semiconductors $M_I M_{II} P_2 X_6$ exhibit the coexistence of two or three types of ferroic orderings such as ferromagnetic, antiferromagnetic, ferroelectric, and ferroelastic orderings, i.e., 2D multiferroics. Specifically, $InHgP_2O_6$ is a multiferroic material with the coexistence of ferroelectricity and ferromagnetism (FE+FM), $NbCuP_2S_6$ and $PtMoP_2O_6$ can accommodate both ferroelectricity and antiferromagnetism (FE+AFM), $GaAuP_2O_6$ is multiferroic with ferroelectricity and ferroelasticity (FE+FEA), while $AlZrP_2O_6$, $HgIrP_2O_6$, and $PtOsP_2O_6$ possess simultaneously three ferroic orderings including ferroelectricity, ferroelasticity, and antiferromagnetism, giving rise to FE+FEA+AFM multiferroics. These multiferroic materials may have potential applications in magnetoelectric, magnetostrictive, or mechanic-electric nanodevices. Besides, it is also interesting to examine their possibilities as photocatalytic materials for water-splitting, as it was reported that ferroelectricity may be beneficial to improve the photocatalytic performance because the built-in electric field hinders the recombination of photogenerated electrons and holes [48,49]. Seven ferroelectric semiconductors, including $ZrZnP_2O_6$, $CdHfP_2O_6$, $GaLaP_2S_6$, $CuTiP_2Se_6$, $CuYP_2S_6$, $CuScP_2S_6$, and $AuScP_2S_6$, were indeed found to have suitable band edges and band gaps (1.59–2.50 eV) for water-splitting photocatalysts (Table S8 online).

*3.4 Ferroelectric metals*

Table 1 lists the formula and properties of 16 ferroelectric metals $M_IM_{II}P_2X_6$, and Fig. S11 (online) presents their geometric structures. They are non-centrosymmetric with space group of *P3* or *P3m1*, corresponding to the "low-symmetry phase" in Fig. 1c, and the corresponding electric polarizations point to the *z*-direction (out-of-plane). As aforementioned, we denoted the absolute values of displacements of $M_I$ and $M_{II}$ atoms with $d_1$ and $d_2$, where the displacements move toward opposite directions for all above ferroelectric metals, leading to the relative vertical displacement between $M_I$ and $M_{II}$ atoms is $d = d_1 + d_2$, as indicated in Fig. 1c. Considering that $M_IM_{II}P_2X_6$ and $M_{II}M_IP_2X_6$ represent the same material, the order of $M_I$ and $M_{II}$ in the formulas in Table 1 is arranged to assure $d_1 > d_2$, which means that $M_I$ contributes primarily to the displacement and polarization for each $M_IM_{II}P_2X_6$. These 16 ferroelectric metals in Table 1 are sorted by polarization (pC/m) in ascending order. We have several interesting observations in order. (i) The ferroelectric metals with high polarization all contain coinage metal elements (Au, Ag, or Cu); (ii) the increasing trend of polarization with $d_1$ has similarity more than with $d_2$ (Fig. S7 online), implying that the displacements of $M_I$ atoms dominate the polarization; (iii) besides Au, Ag, and Cu, $M_I$ elements are In, Ga, Sn or Pb of IIIA or IVA metal elements. By carefully checking the values of $d_1$ ($d_2$) of each element, we can divide the metal elements in ferroelectric metal $M_I M_{II} P_2 X_6$ into three groups: Group A includes coinage metal elements (Au, Ag, or Cu) with large displacements (> 0.62 Å); Group B includes IIIA and IVA metal elements (In, Ga, Sn, and Pb) with moderate displacements (between 0.32 and 0.61 Å); and Group C includes other transition metal elements (Y, Zr, Hf) with small displacements (< 0.28 Å). The coinage metal elements and IIIA/IVA metal elements lead to displacements larger than other transition metal elements. We can understand this behavior as follows. In the high-symmetry (paraelectric) phase of $M_I M_{II} P_2 X_6$, the metal atoms locate in the bisector of P-P pair ($d_1 = d_2 = d = 0$), implying that metal atoms should be bonded with both upper-three and lower-three chalcogen X atoms (Fig. 1a and b), i.e., metal atoms are octahedrally coordinated with six chalcogen atoms, just like that in 1T-$MoS_2$ and other transition metal dichalcogenides (TMDCs) [50]. However, coinage metal elements and IIIA/IVA metal elements have full $d$-orbitals and do not tend to form the octahedrally coordinated bonding. In contrast, they tend to bond only with either upper-three or lower-three chalcogen X atoms, which manifests that the metal atoms will deviate from their original high-symmetric positions, resulting in the spontaneous symmetry breaking and the emergence of the out-of-plane polar axis. Thus, the coinage metal elements and IIIA/IVA metal elements (Group B) with full $d$-orbital, in particular Au, Ag, and Cu, can lead to large displacement and high polarization. It also explains why electronic orbital-based descriptors are essential for high-precision prediction in our machine learning model (Fig. S5c online).

Since the metal atoms in above three groups can have distinct displacements, various combinations of $M_I$ and $M_{II}$ metals may lead to diverse physical properties. Here we focus on $AuZrP_2S_6$ and $InZrP_2Te_6$ as typical examples of Group A+C and Group B+C combinations, respectively. As $d_1$ of $AuZrP_2S_6$ (1.63 Å) is about three times of $InZrP_2Te_6$ (0.52 Å), the polarization of $AuZrP_2S_6$ (9.740 pC/m) is about five times of $InZrP_2Te_6$ (1.822 pC/m). Fig. 3a and b show the energy bands of $InZrP_2Te_6$ and $AuZrP_2S_6$, respectively. The energy bands are almost unaltered when the spin-orbit coupling (SOC) is considered (Figs. S14 and S17 online). For each of InZrP$_2$Te$_6$ and AuZrP$_2$S$_6$, there is a single energy band crossing the Fermi level, indicating a metallic character. This energy band separated from other bands by distinct gaps is exactly half-filling, showing that the number of total valence electrons per unit cell should be odd, as indeed shown in Table 1 for InZrP$_2$Te$_6$ and AuZrP$_2$S$_6$. Other ferroelectric metals exhibit similar characters in energy bands (Figs. S14–S17 online) and electron parity, implying that they share similar electronic properties and the same metallic mechanism. Thus, in all 16 ferroelectric metals we find that a unit cell contains an odd number of electrons, which may lead to a half-filled energy band across the Fermi level, and gives rise to the metallicity. The odd valence electrons in a unit cell appear to be a necessary (but not a sufficient) condition for a ferroelectric metal in M$_I$M$_{II}$P$_2$X$_6$ materials. There are exceptions when the system is spin-polarized however. For instance, InHgP$_2$O$_6$ (Fig. S19a online) has odd valence electrons and its energy band around Fermi level splits into a fulfilled spin-up band and an empty spin-down band with a gap, thus, it is a non-metallic multiferroic with ferroelectric and ferromagnetic orderings. For non-magnetic ferroelectric semiconductors, the number of total valence electrons in a unit cell is even (Table S8 and S9 online), because the even number of electrons in a unit cell would have no unpaired electrons for this family of materials, usually leading to semiconductors.

The projected electronic density of states (PDOS) of AuZrP$_2$S$_6$ and InZrP$_2$Te$_6$ are presented in Fig. 3c and d. For InZrP$_2$Te$_6$, the electronic states at the Fermi level are mainly contributed by the $p$ electrons of Te atoms. Similar electronic structures are also observed in other Group B+C materials (such as PbYP$_2$Te$_6$ and InHfP$_2$Te$_6$, see Figs. S16 and S17 online). In contrast, the electronic states at the Fermi level in AuZrP$_2$S$_6$ are mainly contributed by $d$ electrons of Zr atoms and $p$ electrons of S atoms. Similar electronic structures are also observed in other Group A+C materials (such as AgZrP$_2$S$_6$, CuHfP$_2$Se$_6$, and CuZrP$_2$S$_6$, see Figs. S14 and S15 online). AuSnP$_2$Te$_6$ is a typical Group A+B material, hence there is also a single energy band crossing the Fermi level, but it is mainly contributed by $p$ electrons of Te atoms and $s$ electrons of Sn atoms (Fig. S16 online). Therefore, in Group B+C materials, the $p$ electrons of chalcogen atoms have a dominant contribution to the conducting states; in Group A+C materials, both transition-metal $d$-orbitals and chalcogen $p$-orbitals dominate the conducting states; while in Group A+B materials, the chalcogen *p*-orbitals and *s*-orbitals of IIIA/IVA metal atoms contribute mainly to the conducting states. In all types of above ferroelectric metals, the chalcogen *p*-orbitals play a crucial role in the conduction.

Now we visualize the partial electron densities of InZrP$_2$Te$_6$ and AuZrP$_2$S$_6$ within energy range $|E-E_\text{f}|<$ 0.05 eV, which are usually considered as the conducting electron density $\rho_\text{c}(\vec{r})$ (Details can be found in the Supplementary materials). As shown in Fig. 3e, $\rho_\text{c}(\vec{r})$ of InZrP$_2$Te$_6$ exhibits a *p*-orbital character around Te atoms; in contrast, $\rho_\text{c}(\vec{r})$ of AuZrP$_2$S$_6$ shows a *p-d* hybridization character around Zr and S atoms as indicated in Fig. 3f, both being consistent with the observations from PDOS. The conducting electrons of InZrP$_2$Te$_6$ and AuZrP$_2$S$_6$ constitute a C3-symmetry connecting network with a few low-density hollows, which may provide conducting channels in real space. The left panels of Fig. 3g and h show the side views of an isosurface of $\rho_\text{c}(\vec{r})$. It is surprising to observe that $\rho_\text{c}(\vec{r})$ of InZrP$_2$Te$_6$ and AuZrP$_2$S$_6$ are mainly distributed on the upper surface of the 2D layer and are weakly relevant to In and Au atoms (indicated with blue color) of the lower surface. Note that In and Au are M$_\text{I}$ atoms, which have large displacements and are major contributors for the electric polarization. We define a 'reduced' conducting electron density by integrating $\rho_\text{c}(\vec{r})$ over the *x-y* plane, say, $\rho_\text{c}(z) = \iint \rho_\text{c}(\vec{r})\text{d}x\text{d}y$, where $\rho_\text{c}(\vec{r})$ is the conducting electron density. The results are shown in the right panels of Fig. 3g and h, which indicate clearly that the conducting electrons mainly move on the upper surface. When M$_\text{I}$ atoms move to the lower surface, the coordinate number for chalcogen atoms on the upper surface could be reduced and excessive electrons emerge, which may partially fill the energy band around Fermi level and contribute to the conduction.

To describe the spatial distribution of electrons that contribute to the electric polarization, we introduce a "reduced" difference charge density defined as $\rho_\text{p}(z) = \iint [\rho_\text{FE}(\vec{r}) - \rho_\text{PE}(\vec{r})]\text{d}x\text{d}y$, where $\rho_\text{FE}(\vec{r})$ and $\rho_\text{PE}(\vec{r})$ are the total electron densities of a ferroelectric material in ferroelectric and paraelectric phases, respectively. Since the paraelectric phase has a high symmetric structure and zero polarization, $\rho_\text{PE}(\vec{r})$ has no contribution to polarization. The difference between $\rho_\text{FE}(\vec{r})$ and $\rho_\text{PE}(\vec{r})$ can reflect the electronic contribution to the polarization, and we term $\rho_\text{p}(z)$ as the reduced "FE-PE electron density difference". As shown in the right panels of Fig. 3g and h, $\rho_\text{p}(z)$ exhibits an oscillating behavior and reveals the spatial distribution of charge polarization. $\rho_p(z)$ of AuZrP$_2$S$_6$ is closer to the lower surface than that of InZrP$_2$Te$_6$, which is consistent with the fact that the displacement of Au atoms is larger than that of In atoms, and Au atoms are also closer to the lower surface of the 2D layer. For both materials $\rho_c(z)$ deviates obviously from that of $\rho_p(z)$, i.e., the conducting electrons and "FE-PE electrons density difference" are spatially separated, providing clues on the underlying mechanism of ferroelectric metallicity in these 2D M$_I$M$_{II}$P$_2$X$_6$ metals.

The coexistence of ferroelectricity and metallicity in 2D M$_I$M$_{II}$P$_2$X$_6$ materials can be rationalized as follows. First, the chemical nature of coinage metal elements and IIIA/IVA metal elements (M$_I$ site) make them deviate from high-symmetric positions and move to the lower surface (Fig. 3g and h), which have large displacements and make the major ionic contribution to electric polarization. In contrast, the displacements of M$_{II}$ atoms are tiny, resulting in a rather weak effect. Along with atomic displacements, the "FE-PE electron density difference" distributes mainly in the lower part of the 2D layer. Thus, the total polarization including ionic and electronic contributions is related to the lower part of 2D M$_I$M$_{II}$P$_2$X$_6$ materials. Second, when the total number of valence electrons in a unit cell is odd, and the system is not spin-polarized, the energy band around Fermi level should be half-filled, leading to a metallic property. M$_I$ atoms move to the lower surface and leave unsaturated chalcogen atoms and excessive electrons on the upper surface, which contribute to the conduction. Third, the conducting electron density dominantly distributes around the chalcogen atoms and M$_{II}$ atoms in the upper surface, which cannot completely screen the vertical polarization that mainly comes from the lower part of the 2D layer. In addition, the low-density hollows are observed in conducting electron density (Fig. 3e and f), which implies that the conducting electrons may not completely exclude the external electric fields. Our analyses reveal that the coexistence of ferroelectricity and metallicity here is the consequence of a dimensionality effect. These findings suggest that ferroelectric metals could be highly achievable in 2D materials.

### *3.5 Ferroelectric-paraelectric phase transition and polarization reversal*

The phase transition between ferroelectric and paraelectric phases is an essential character of ferroelectric materials and crucial for possible applications. We simulated the polarization reversal paths of these 16 ferroelectric metals $M_IM_{II}P_2X_6$ by using the climbing image nudged elastic band (CI-NEB) method [51]. The ferroelectric-paraelectric transition barriers are obtained, as listed in Table 1, which range from 0.04 to 0.57 eV/unit cell. The barriers generally increase with the increase of displacement of $M_{II}$ atoms ($d_2$) (see Fig. S8b online). For two ferroelectric metals, $AuHfP_2O_6$, and $AuZrP_2O_6$, as shown in Fig. 4a and b, the energy versus polarization profiles exhibit common double-well shape and clear bistability, where two minima correspond to ferroelectric phases of opposite polarizations, while the maximum corresponds to paraelectric phase (see Fig. 4a, b for energy versus polarization profiles of $AuHfP_2O_6$, and $AuZrP_2O_6$). The Landau-Ginzburg theory was applied to deal with ferroelectric phase transitions. We take polarization ***P*** as the order parameter of ferroelectric phase. The Landau-Ginzburg expansion has the form of $E = \sum_i \left[\frac{A}{2}P_i^2 + \frac{B}{4}P_i^4 + \frac{C}{6}P_i^6\right] + \frac{D}{2}\sum_{<i,j>}(P_i - P_j)^2$. Here $P_i$ is the polarization of *i*-th unit cell, <i,j> indicates the nearest neighbors, and *A, B, C, D* are coefficients. The energy versus polarization profiles in Fig. 4 is fitted using the first three terms. The last term describes the dipole-dipole interaction between nearest neighboring unit cells. All the coefficients are obtained by fitting to DFT results, as listed in Table S11 (online). Based on this model, ferroelectric Curie temperatures $T_c$ are obtained using Monte Carlo (MC) simulation, which was frequently employed to estimate ferroelectric transition temperature $T_c$ in previous works [52,53]. It is predicted that $T_c$ of $AuZrP_2O_6$ and $AuHfP_2O_6$ is 2050 and 800 K (Fig. S25 online), respectively, which are much higher than room temperature, showing that the ferroelectricity of these metals is very robust under thermal perturbations. The ultrahigh $T_c$ of $AuZrP_2O_6$ may be due to strong dipole-dipole interaction (*D*) between nearest neighboring unit cells.

As aforementioned, bilayer $WTe_2$ is the first experimental evidence for the coexistence of ferroelectricity and metallicity in 2D materials, which has a polarization of 0.375 pC/m ($2.00 \times 10^{11}$ e/cm$^2$) at 20 K [23]. Our calculated polarization for bilayer $WTe_2$ is 0.423 pC/m ($2.35 \times 10^{11}$ e/cm$^2$), in good agreement with the experimental value, indicating our calculations are reliable. As listed in Table 1, we find that the polarization of 16 ferroelectric metals $M_IM_{II}P_2X_6$ is about one order of magnitude higher than bilayer $WTe_2$. Besides, the Curie temperatures are also considerably higher than that of bilayer $WTe_2$. Thus, the robustness of metallic ferroelectricity in $M_IM_{II}P_2X_6$ is stronger than that in bilayer $WTe_2$, implying that the coexistence of ferroelectricity and metallicity in $M_IM_{II}P_2X_6$ may be easily experimentally detected than in bilayer $WTe_2$.

Generally, the polarization of a ferroelectric material can be switched by applying a proper electric field, which was successfully demonstrated in bilayer $WTe_2$ [23]. We further calculated the energy versus polarization profiles under different vertical electric fields for $AuHfP_2O_6$ and $AuZrP_2O_6$. As shown in Fig. 4c and d, one may see that with increasing the electric field, the energies increase for P↓ polarization and decrease for P↑ polarization, which makes the energy barrier from P↓ to P↑ polarization decrease dramatically. It turns out that an electric field of 1.0 V/Å gives rise to small energy barriers of 10 and 26 meV/unit cell for $AuHfP_2O_6$ and $AuZrP_2O_6$, respectively, which suggests that 1.0 V/Å may be the critical electric field for reversing polarization at room temperature. Note that the structures cannot be destroyed until the electric field is higher than 2.8 V/Å. This critical electric field (1.0 V/Å) is nearly the same order of magnitude as that of 2D ferroelectric $In_2Se_3$ (0.66 V/Å) [15]. As the electric polarization of the latter can be switched by gate voltage experimentally [54], the polarization of $AuHfP_2O_6$ and $AuZrP_2O_6$ could likewise be switched by a proper external electric field.

*3.6 Van der Waals heterostructures based on ferroelectric metal*

The coexistence of ferroelectricity and metallicity in $M_IM_{II}P_2X_6$ may bring various potential applications. The most intriguing property of 2D ferroelectric metal $M_IM_{II}P_2X_6$ is the fact that conducting electrons distribute mainly on single side of 2D layer with finite thickness, which is associated with the polarization direction. If we flip the direction of polarization up or down, major conducting electrons of $M_IM_{II}P_2X_6$ will move to the lower or upper surface of the 2D layer, which may lead to novel physical effects. The van der Waals contact between 2D ferroelectric metal and other 2D materials would be appropriate to demonstrate how the switching of polarization direction modulates the surface or interface properties of the contact. By considering lattice matching, we constructed two types of van der Waals heterostructures as examples, i.e., AuSnP$_2$Te$_6$/graphene and AuZrP$_2$S$_6$/MoS$_2$. Fig. 5a and b show the energy bands of AuSnP$_2$Te$_6$/graphene heterostructure with the polarization of AuSnP$_2$Te$_6$ along $z$ direction (P↑) and opposite direction (P↓), respectively. The energy bands from graphene and AuSnP$_2$Te$_6$ shift up and down relative to the Fermi level, respectively, indicating a transfer of electrons from graphene to AuSnP$_2$Te$_6$ (Fig. S27f online), while the profiles of energy bands remain intact. The shift of energy bands for graphene with polarization P↑ is larger than that for P↓, showing that the reversal of polarization of AuSnP$_2$Te$_6$ would influence the number of conducting electrons transferred (Fig. S27h online). Thus, in this heterostructure of graphene and AuSnP$_2$Te$_6$ in which both are conducting, an electric field may induce different resistive states depending on the direction of electric polarization, which may have potential applications in information storage.

Fig. 5c and d show the energy bands of AuZrP$_2$S$_6$/MoS$_2$ heterostructure with the polarization of AuZrP$_2$S$_6$ along $z$-direction (P↑) and opposite direction (P↓), respectively. It is interesting to note that there is still a half-filled band across Fermi level, which comes from AuZrP$_2$S$_6$ and locates in the gap of MoS$_2$, indicating that no charge transfers between AuZrP$_2$S$_6$ and MoS$_2$ layers, which is caused by the large gap of MoS$_2$ (~1.7 eV). Regardless of the polarization direction of AuZrP$_2$S$_6$, the Fermi level is not obviously shifted from the half-filled band, however, the energy bands from MoS$_2$ shift remarkably when the direction of polarization is switched from P↑ to P↓. The conduction band minima (CBM) of MoS$_2$ is 0.708 eV above Fermi level in AuZrP$_2$S$_6$ (P↑)/MoS$_2$ and 0.375 eV above Fermi level in AuZrP$_2$S$_6$ (P↓)/MoS$_2$ (Fig. 5c and d), which can be viewed as the Schottky barrier ($\Phi_e$) between AuZrP$_2$S$_6$ and intrinsic MoS$_2$. Thus, the contact between AuZrP$_2$S$_6$ and intrinsic MoS$_2$ is of Schottky type, where the Schottky barrier can be modulated through the reversal of electric polarization. In addition, doping in semiconductor MoS$_2$ can induce an impurity level in the gap. If the impurity level ($E_{im}$) is below the CBM and the value of $E_{CBM}-E_{im}$ is between 0.708 and 0.375 eV, for example, fluorine doping in MoS$_2$ will induce an impurity level with $E_{CBM}-E_{im}$ = 0.5 eV [55], the contact of AuZrP$_2$S$_6$ (P↑)/MoS$_2$ will still be of Schottky type, while the contact of AuZrP$_2$S$_6$ (P↓)/MoS$_2$ will become Ohmic type. In other words, the reversal of electric polarization can switch the contact from Schottky to Ohmic type for the heterostructure of AuZrP$_2$S$_6$ and properly doped MoS$_2$. The change of the height of Schottky barrier or the Schottky-Ohmic contact type could induce a dramatic change of the transport properties along the vertical direction, and novel physical effects such as giant electroresistance [56,57] could be expected.

By combining 2D ferroelectric metals and other 2D materials, various types of van der Waals heterostructures can be generated, and their physical properties can be easily modulated through doping, an electric field or a magnetic field, etc., to make them possible potential candidates for novel devices such as ferroelectric tunneling junction with giant electroresistance, field-effect transistor, nonvolatile ferroelectric memory, etc. Besides, considering the coupling between the direction of polarization and single-side surface conducting electrons, ferroelectric metals M$_I$M$_{II}$P$_2$X$_6$ may be good materials for the use in electrical-writing/optical-reading memory and ferroelectric printing devices.

## 4. Conclusion

We employed high-throughput *ab initio* calculations and data-driven machine learning schemes with a set of new electronic orbital-based descriptors to discover novel 60 stable ferroelectric materials out of 2,964 structures of 2D bimetal phosphates M$_I$M$_{II}$P$_2$X$_6$, including 16 ferroelectric metals with out-of-plane electric polarization and 44 ferroelectric semiconductors. Among the 44 semiconducting ones, seven multiferroics with two or three types of ferroic orderings and seven ferroelectric water-splitting photocatalysts were screened out. For this family of 2D ferroelectric metals, the physical origin of ferroelectricity is owing to the spontaneous symmetry breaking induced by the opposite vertical displacements of two metal atoms M$_I$ and M$_{II}$, where the out-of-plane polarization (***P***) and relative vertical displacement (*d*) are unveiled to comply a quadratic relationship, revealing that these 16 2D ferroelectric metals belong to displacive-type ferroelectrics. The coinage metal elements with full *d*-orbitals, i.e., Au, Ag, and Cu, can lead to large displacements and large polarization due to the nature of their chemical bonding. We also found that odd number of valence electrons in a unit cell lead to a half-filled energy band around Fermi level and is responsible for metallicity, and chalcogen *p*-orbitals play an important role in the conduction in all 16 ferroelectric metals. The charge density analysis shows that the coexistence of ferroelectricity and metallicity in $M_IM_{II}P_2X_6$ is a consequence of dimensionality effect, in which the conducting electrons distribute mainly on the upper surface of the 2D layer, whereas the out-of-plane electric polarization induced from both ionic and electronic motion is primarily related to the lower part of the 2D layer. It shows if conducting electrons move on a 2D plane, while the electric polarization happens along the third dimension, ferroelectricity and metallicity could coexist in the same material. With this scenario, ferroelectric metals could be achievable in 2D materials. Our work presents a large family of novel 2D ferroelectric metals with intriguing properties, which could be applied to construct heterostructures through van der Waals contact that may have possible applications in areas of ferroelectric tunneling junction, nonvolatile ferroelectric memory, etc. As ferroelectric metals are still extremely sparse now, it would spur great interest in exploring 2D ferroelectric metals in near future.

## Conflict of interest

The authors declare that they have no conflict of interest.

## Acknowledgments

This work was supported in part by the National Key R&D Program of China (2018YFA0305800), the Strategic Priority Research Program of CAS (XDB28000000), the National Natural Science Foundation of China (11834014), Beijing Municipal Science and Technology Commission (Z191100007219013), and University of Chinese Academy of Sciences. The calculations were performed on Era at the Supercomputing Center of the Chinese Academy of Sciences and Tianhe-2 at the National Supercomputing Center in Guangzhou.

## Author contributions

Qing-Bo Yan and Gang Su conceived the project and supervised the research. Xing-Yu Ma performed the calculations. Xing-Yu Ma, Qing-Bo Yan, and Gang Su developed the codes for calculating out-of-plane polarization and machine learning model, analyzed the results, and wrote the manuscript. Xiaofeng Qian participated in data analysis and manuscript writing. Hou-Yi Lyu, Kuan-Rong Hao, and Yi-Ming Zhao participated in the discussion.

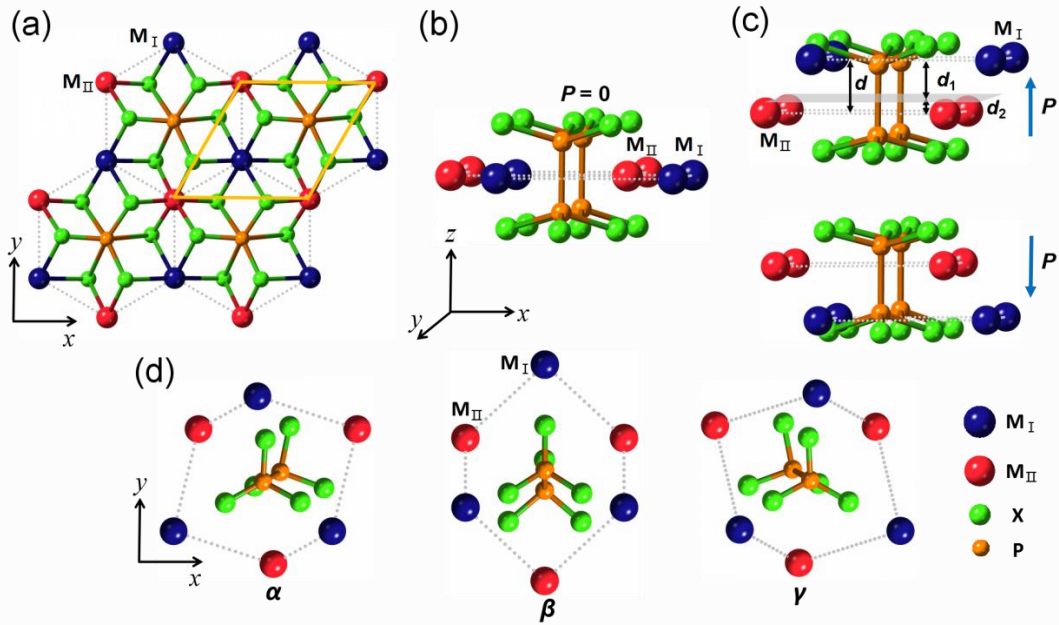

**Fig. 1.** Schematic structures of 2D bimetal phosphates ($M_IM_{II}P_2X_6$, $M_I$ and $M_{II}$ atoms are different metal elements, X=O, S, Se, Te). The blue, red, green and orange balls represent $M_I$ metal atoms, $M_{II}$ metal atoms, chalcogen atoms, and phosphorus atoms, respectively. (a) Top view. The yellow parallelogram indicates the unit cell and dash lines denote a honeycomb lattice formed by metal atoms. P atoms sit at the center of the hexagons, and chalcogen atoms bridge metal atoms and P atoms. (b) Side view of the high-symmetry phase. (c) Side view of Type-I low-symmetry phases. $d$ is the vertical displacement between $M_I$ and $M_{II}$ atoms. $d_1$ and $d_2$, are the displacements of $M_I$ and $M_{II}$ atoms related to the bisect plane (indicated with grey color) of P-P pairs. The blue arrows (up/down) indicate the directions of out-of-plane electric polarization (***P***). (d) Top view of Type-II low-symmetry phases. The distorted structure with P-P pairs inclining along different directions forms three symmetry-equivalent phases (*α*, *β* and *γ*), respectively.

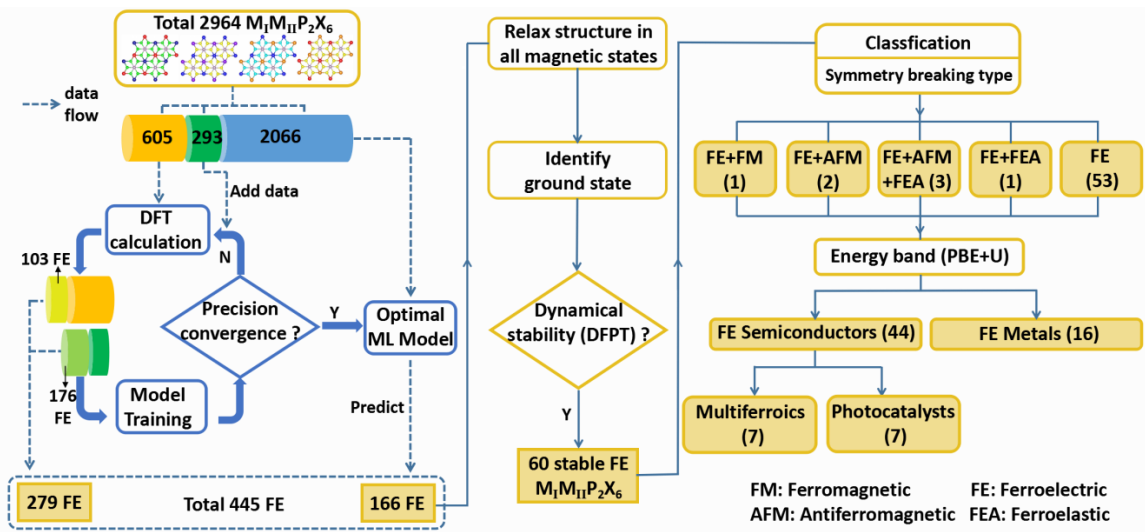

**Fig. 2.** The schematic procedure for discovering ferroelectric materials $M_IM_{II}P_2X_6$. The left panel is the flowchart of the data-driven machine learning process, which generates an optimal machine learning model and obtains 445 potential ferroelectrics. In the right panel, 60 stable ferroelectric materials $M_IM_{II}P_2X_6$ are obtained after dynamically stability screening, which are further classified into different types of ferroelectrics, and finally, 16 ferroelectric metals are figured out.

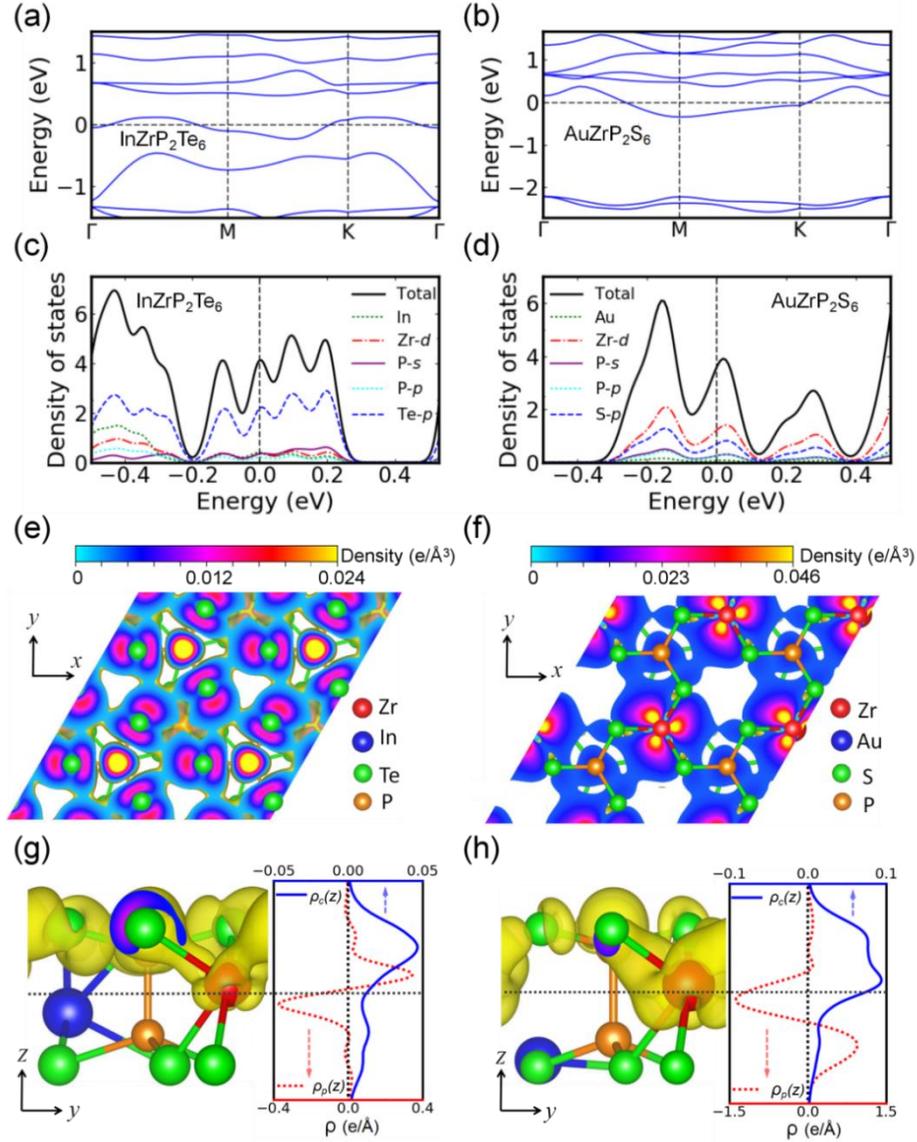

**Fig. 3** Electronic properties of two ferroelectric metals: InZrP$_2$Te$_6$ and AuZrP$_2$S$_6$. (a, b) Band structures at PBE+U level. (c, d) Projected density of states (PDOS). (e, f) and left panels of (g, h) are top and side views of partial electron density within energy range $|E-E_f| < 0.05$ eV, i.e., $\rho_c(\vec{r})$, respectively. In right panels of (g, h), blue lines indicate the reduced conducting electron density $\rho_c(z)$, and red dash lines indicate the reduced "FE-PE electron density difference" $\rho_p(z)$, where the vertical axis denotes the spatial $z$ coordinate that is the same as in geometrical structures of corresponding left panels.

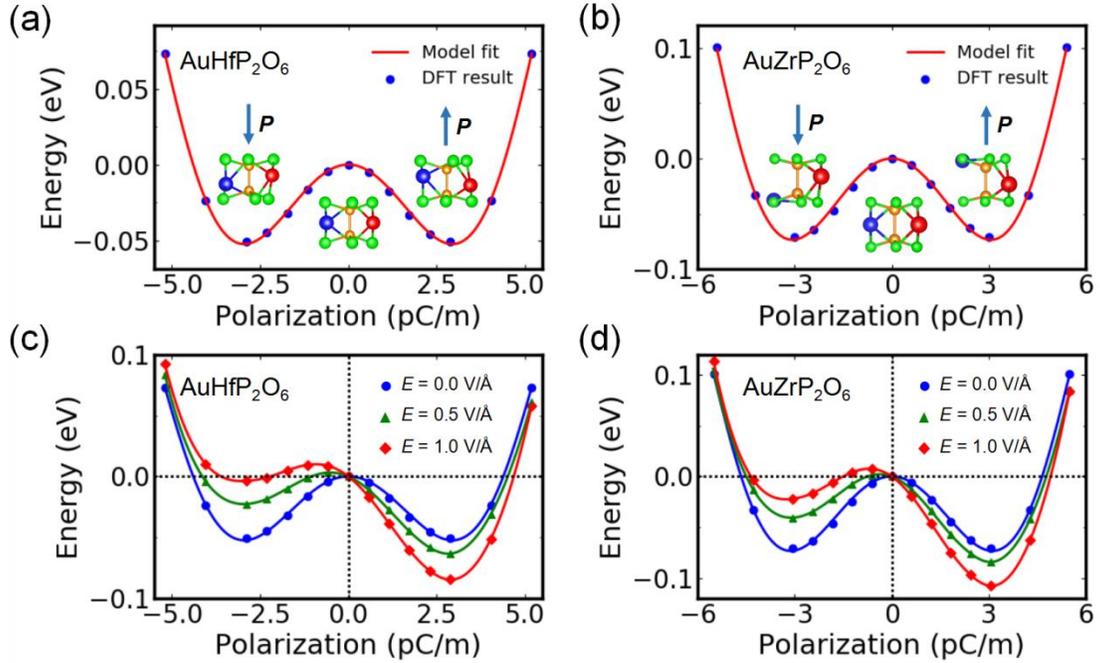

**Fig. 4.** Energy versus polarization of two ferroelectric metals (AuHfP$_2$O$_6$ and AuZrP$_2$O$_6$) and the effects of external electric field. (a, b) Energy versus polarization of two ferroelectric metals. The barrier between ferroelectric and paraelectric phases is presented. Blue points are the DFT-calculated total energies. The red lines are fitted curves with Landau-Ginzburg model. The directions of polarization (***P***) in ferroelectric phases are marked by blue arrows. (c, d) Energy versus polarization of two ferroelectric metals under different vertical external electric fields. Points are the DFT-calculated total energies, and the lines are fitted curves.

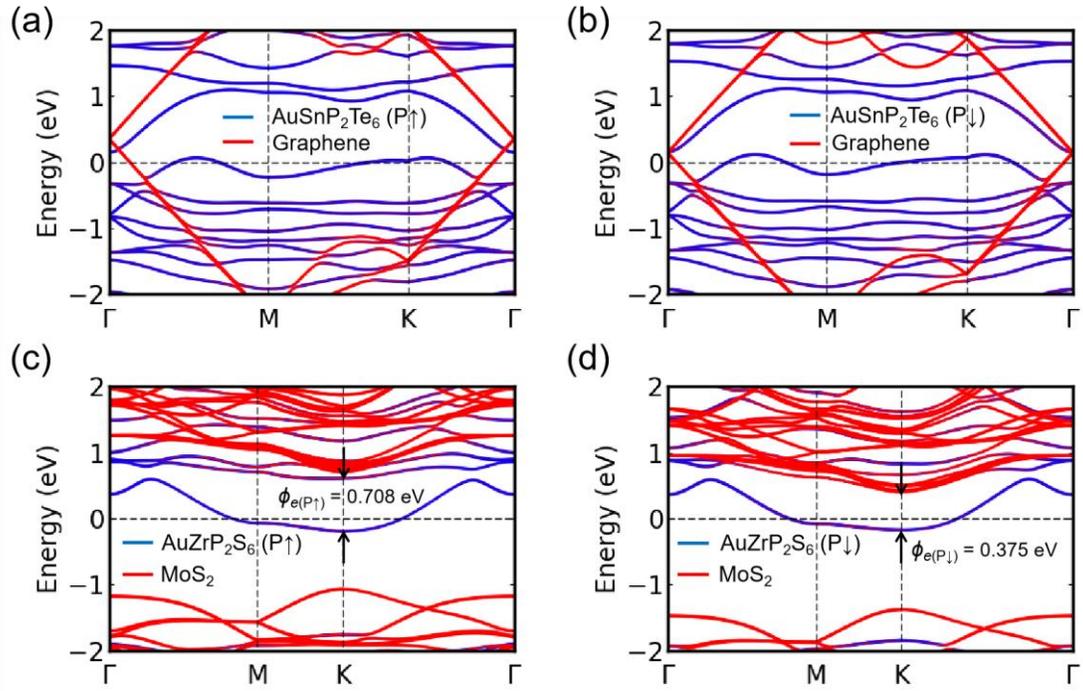

**Fig. 5.** Band structures of heterostructures. (a, b) $AuSnP_2Te_6$/Graphene. (c, d) $AuZrP_2S_6$/$MoS_2$. The directions of polarization (*P*) in ferroelectric metals are marked by black thin arrows. The blue and red colors indicate the contributions from different 2D materials.

**Table 1** Properties of 16 ferroelectric metals.

| Formula | Space group | $d$ (Å) | $d_1$ (Å) | $d_2$ (Å) | Polarization (pC/m)/ (eÅ/ unit cell) | Barrier (eV/unit cell) | $N_e$ / unit cell |
|---|---|---|---|---|---|---|---|
| $InPbP_2Te_6$ | P3 | 1.16 | 0.61 | 0.55 | 0.570 (0.017) | 0.360 | 53 |
| $SnYP_2Te_6$ | P3 | 0.67 | 0.42 | 0.25 | 0.840 (0.025) | 0.169 | 53 |
| $GaSnP_2Te_6$ | P3 | 0.88 | 0.47 | 0.41 | 1.009 (0.028) | 0.249 | 53 |
| $GaZrP_2Te_6$ | P3 | 0.66 | 0.42 | 0.24 | 1.191 (0.034) | 0.145 | 53 |
| $GaHfP_2Te_6$ | P3 | 0.58 | 0.40 | 0.18 | 1.353 (0.038) | 0.148 | 53 |
| $InSnP_2Te_6$ | P3 | 1.08 | 0.61 | 0.47 | 1.610 (0.046) | 0.420 | 53 |
| $InZrP_2Te_6$ | P3 | 0.80 | 0.52 | 0.28 | 1.822 (0.053) | 0.195 | 53 |
| $InHfP_2Te_6$ | P3 | 0.71 | 0.49 | 0.22 | 1.916 (0.056) | 0.187 | 53 |
| $PbYP_2Te_6$ | P3 | 0.72 | 0.47 | 0.25 | 2.131 (0.064) | 0.175 | 53 |
| $AuHfP_2O_6$ | P3m1 | 0.67 | 0.62 | 0.05 | 2.888 (0.045) | 0.044 | 61 |
| $AuZrP_2O_6$ | P3m1 | 0.74 | 0.69 | 0.05 | 3.009 (0.047) | 0.063 | 61 |
| $AgZrP_2S_6$ | P3 | 0.94 | 0.78 | 0.16 | 3.010 (0.064) | 0.096 | 61 |
| $CuHfP_2Se_6$ | P3 | 1.47 | 1.36 | 0.11 | 5.463 (0.123) | 0.146 | 61 |
| $CuZrP_2S_6$ | P3 | 1.54 | 1.39 | 0.15 | 5.884 (0.121) | 0.182 | 61 |
| $AuZrP_2S_6$ | P3 | 1.83 | 1.63 | 0.20 | 9.740 (0.206) | 0.573 | 61 |
| $AuSnP_2Te_6$ | P3 | 2.09 | 1.77 | 0.32 | 9.828 (0.269) | 0.365 | 61 |
| Bilayer $WTe_2$ | Pm | - | - | - | 0.423 (0.0058) | 0.0006 | - |

Note: $d$ is the vertical relative-displacement between $M_I$ and $M_{II}$ metal atoms along $z$-direction as indicated in Fig. 1c. $d_1$ and $d_2$ are the absolute values of displacements of $M_I$ and $M_{II}$ atoms when taking the bisecting plane of *P-P* pair as a reference, and $d_1 + d_2 = d$. The polarizations are listed in two different units. The ferroelectric-paraelectric transition barriers and number of total valence electrons ($N_e$) are also listed. The properties of bilayer $WTe_2$ are listed for comparison, which are obtained with the same method.